\documentclass[sigconf]{acmart}

\usepackage{enumitem}

\copyrightyear{2026}
\acmYear{2026}
\setcopyright{cc}
\setcctype{by}
\acmConference[IDE '26]{3rd International Workshop on Integrated Development Environments }{April 12--18, 2026}{Rio de Janeiro, Brazil}
\acmBooktitle{3rd International Workshop on Integrated Development Environments (IDE '26), April 12--18, 2026, Rio de Janeiro, Brazil}
\acmDOI{10.1145/3786151.3788606}
\acmISBN{979-8-4007-2384-1/2026/04}

\begin{document}

\title{Detecting UX smells in Visual Studio Code using LLMs}

\author{Andrés Rodriguez}
\email{arodrig@lifia.info.unlp.edu.ar}
\affiliation{%
  \institution{LIFIA, Fac. Informática, \\Univ. Nac. La Plata}
  \city{La Plata}
  \country{Argentina}
}

\author{Juan Cruz Gardey}
\email{jcgardey@lifia.info.unlp.edu.ar}
\affiliation{%
  \institution{LIFIA, Fac. Informática, \\Univ. Nac. La Plata}
  \city{La Plata}
  \country{Argentina}}

\author{Alejandra Garrido}
\email{garrido@lifia.info.unlp.edu.ar}
  \affiliation{%
      \institution{LIFIA, Fac. Informática, \\Univ. Nac. La Plata \&  CONICET}
      \city{La Plata}
      \country{Argentina}
    }

\renewcommand{\shortauthors}{Rodriguez et al.}

\begin{abstract}
Integrated Development Environments shape developers’ daily experience, yet the empirical study of their usability and user experience (UX) remains limited. This work presents an LLM-assisted approach to detecting UX smells in Visual Studio Code by mining and classifying user-reported issues from the GitHub repository. Using a validated taxonomy and expert review, we identified recurring UX problems that affect the developer experience. Our results show that the majority of UX smells are concentrated in informativeness, clarity, intuitiveness, and efficiency, qualities that developers value most.   
\end{abstract}



\maketitle

\section{Introduction}

Integrated Development Environments (IDEs) play a central role in shaping developers’ everyday experience with code. Far from being neutral instruments, IDEs mediate cognition, workflow, and collaboration, influencing how developers search, refactor, and reason about software \cite{fagerholm2012developer}. Over the past decade, the scope and complexity of IDEs have increased dramatically;  modern platforms such as Visual Studio Code (VSCode), IntelliJ IDEA and Eclipse, integrate not only editors and compilers but also live collaboration tools, AI-assisted completion, and plugin ecosystems that redefine how developers interact with their codebases. This evolution has made the developer experience (DEX) an essential quality dimension of software tools \cite{fagerholm2012developer}.

Despite this shift, the empirical study of IDE usability and developer experience remains fragmented. Much of the literature still focuses on feature-level performance or adoption metrics, rather than on the nuanced forms of interaction friction that developers face in daily use. Kuusinen \cite{Kuusinen2015} observed that developers appreciate IDEs that are efficient to use, flexible, informative and intuitive. These qualities, while generally understood, are rarely used as an analytical framework to assess or monitor the user experience (UX) health of an IDE over time.

Parallel to this, recent software engineering research has drawn attention to the notion of UXDebt: a form of debt that accumulates when UX issues are postponed or insufficiently addressed during development \cite{rodriguez2023ux,baltes2024ux}. 
In complex tools such as IDEs, this debt often materializes as UX smells: recurring patterns of interaction breakdowns, confusing feedback, or inconsistency between user expectations and system behavior 
\cite{grigera2017automatic}.
Identifying such UX smells in real-world development tools remains challenging. Controlled usability testing is rarely feasible for open, continuously evolving IDEs with millions of users.  

In this paper, our goal is to mine issues from public repositories, such as GitHub, for evidence of UX smells directly from user reports and community dialogue. 
The volume and diversity of user reports available in large open repositories allow observing how developers articulate friction points, how maintainers triage them, and how recurring UX problems persist or evolve.
Yet, the scale of data volumes requires automating the analysis. 
Thus, we introduce a Large Language Model (LLM)
assisted mining approach that leverages recent advances in natural language understanding to act as a first-pass semantic coder over developer discourse. Using  LLMs, we support the semantic categorization of UX smells according to the 
desirable IDE qualities identified by \cite{Kuusinen2015}. 
The combination of manual inspection and LLM-assisted classification aims to balance interpretive depth with scalability, integrating automation with human judgment.

Our contributions are: (1) an empirical corpus of IDEs’ UX smells grounded in developer discourse, 
rising the understanding of the human aspects of software engineering, 
and (2) foundations for a broader characterization of UXDebt in IDEs, connecting design-level frictions with potential downstream consequences, such as cognitive overload, inefficient workflows, or even the accumulation of Technical Debt/UXDebt in the resulting code.

\section{Background on UX, UX Smells and UXDebt}

UX is an essential aspect of a product, determining its quality as well as its success. The notion of UX considers not only the pragmatic aspects of interaction (functionality, interactive behavior,
user skills, context of use) but also the hedonic (brand image, presentation, internal state of the user resulting from previous experiences) \cite{hassenzahl2021user}.

UX evaluation is often neglected, especially in agile methods that are driven by customer satisfaction and short iteration cycles. Therefore, lightweight methods are needed to evaluate UX as part of iterative development. One method previously proposed is UX refactoring, defined as changes applied with the purpose of improving UX quality while preserving functionality \cite{grigera2017automatic}. In turn, a UX smell hints at a problem with the navigation, presentation, interaction, or any UX aspect that may be solved by applying UX refactoring. An example of a UX smell is a free text input that only accepts a small set of possible values ("Free input for limited values"); it may be solved by applying alternative UX refactorings like "Add Autocomplete" or "Turn Input into Select". Another type of UX smell refers to issues related to the style or aesthetics of a user interface (UI) such as low color contrast, misaligned elements, and lack of responsiveness, among others. Note that UX smells are different from bugs in the UI since smells do not prevent users from accomplishing their goal but just make it cumbersome or uncomfortable.
The presence of UX smells may contribute to the accumulation of UXDebt. This concept has been defined as a type of Technical Debt (TD) with a cumulative cost for the development team as well as stakeholders \cite{rodriguez2023ux}. Similar to TD, UXDebt can undermine code maintainability and increase rework costs.  

\section{Related Work on DEX}
There are several studies that evaluate the usability and UX of IDEs through empirical and/or inspection methods \cite{kline2005evaluation}. Moreover, Fagerholm and Munch defined the concept of Develop Experience (DEX), to help understand developers' perceptions and feelings as users of IDEs, and with the assumption that improving DEX will have a positive impact on productivity \cite{fagerholm2012developer}. They define DEX as consisting of three dimensions: cognitive (perceptions of  development infrastructure), affective (feelings about work and social aspects), and conative (alignment of developers and project goals). 

Further studies on DEX tried to gain an understanding of developers' expectations in the use of IDEs, through coding camps \cite{palviainen2015design} and surveys \cite{Kuusinen2015}.
In the first case, authors study online collaborative coding and categorize IDE features supporting DEX at the level of operations, actions and activities \cite{palviainen2015design}. 
Moreover, they highlight that DEX is composed not only of the experience of using tools, but also the processes involved, the rules, and other people.
In the second case, Kuusinen asked developers about the best qualities of an IDE and the improvements that could better support their work \cite{Kuusinen2015}. The author found that
developers expect IDEs to be more flexible, informative, and reliable. 

There are two works that specifically study the Visual Studio IDE. Amann et al. present a large empirical study with C\# developers on how they use their time in the IDE, although they do not report on its usability \cite{amann2016study}. 
In the study from Vaithilingam et al., the authors conducted user tests with 61 programmers at Microsoft, over several prototype interfaces for change suggestions in VSCode \cite{vaithilingam2023towards}. 
Through a user study they found a better design that improved the usage of change suggestions. 

Thus, previous studies report on user tests or manual expert inspection, which are usually limited in volume and costly, while our approach of analyzing DEX in issue repositories is extensive, automated and low-cost.

\section{Method}

We adopted a mixed approach combining repository mining, LLM-assisted text analysis, and expert validation to identify and characterize UX smells in the VSCode project. Our methodological goal was to balance scalability enabled by automation, with interpretive reliability ensured through human judgment and iteration. The process included three stages described below: 
(1) data collection, (2) LLM-assisted categorization, and (3) synthesis and interpretation.

\subsection{Data Collection}

Repository mining is a common strategy for investigating user experience in real-world settings, allowing researchers to capture "naturally occurring evidence" of interaction breakdowns and user perceptions at scale \cite{Panichella2015MiningDevFeedback}.
We extracted all issues from the public GitHub repository of VSCode\footnote{\url{https://github.com/microsoft/vscode}} using the GitHub REST API. From this corpus, we retained only those issues explicitly tagged with the label UX, a convention used by the maintainers to flag UX-related reports. This filtering step provided an initial dataset of N = 2350 issues (as of October 2025)\footnote{The complete data set is available at \url{https://t.ly/dRhwP}}.

\subsection{LLM-Assisted Categorization of UX Smells}

The filtered issues were processed using an LLM to support semantic categorization regarding an existing catalog of UX smells \cite{grigera2017automatic}. 
The prompt instructed the model to: (i) identify UX-related problems that match entries in the UX smell catalog; (ii) associate each issue with the most relevant UX smell, providing a short rationale describing the reasoning behind the classification; and (iii) detect any additional UX smells not represented in the catalog, propose their tentative label, and link them to the corresponding issue(s) with justification. 
This LLM-assisted annotation follows emerging methodological practices in software repository mining, where models serve as first-pass semantic coders to reveal latent structures in unstructured developer discourse \cite{Abedu2024LLMChatbotsMSR}.
All LLM-assisted classifications were conducted with OpenAI GPT-5, accessed via the ChatGPT interface (April 2025 build).

To ensure validity and reliability, three researchers independently reviewed the same elements from 2 sample sets of LLM output: 
(i) a 10\% random sample of issues classified by the LLM, verifying correctness and rationale clarity, and (ii) a 10\% sample of unclassified issues, checking for missed or ambiguous cases.
Disagreements and misclassifications were discussed in consensus meetings following a constant comparison approach inspired by grounded theory methodology \cite{Charmaz2014ConstructingGroundedTheory}, leading to the construction of a validated label for 236 manually reviewed issues. This set yielded a baseline accuracy of 0.695 between the LLM-assigned and validated labels. Subsequently, we conducted a calibration phase using an updated prompt and an extended catalog. A confusion matrix and per-category metrics revealed systematic errors informing heuristic adjustments and a full reclassification. To preserve rigor, we adopted a hybrid strategy: human-validated labels take precedence, while calibrated heuristics cover unreviewed issues. A second manual inspection (more than 80\% of confirmations) validated this strategy. This semi-supervised approach broadens corpus coverage and forms the empirical basis for subsequent analyses \cite{liu2017improving}.

\subsection{Descriptive and Interpretive Analysis}
\label{sec:interpretiveAnalysis}

In the third phase, we conducted: (i) \textbf{Descriptive statistics}, quantifying the frequency and distribution of UX smells across the dataset, and (ii) \textbf{Analytical mapping}, 
relating UX smells with IDE qualities \cite{Kuusinen2015}, followed by the assignment of IDE qualities to issues. Step (ii) was also LLM-assisted and manually reviewed, allowing us to explore how different forms of UX friction reflected in the issues, cluster around specific experiential qualities valued by developers.

The interpretive stage aims to connect patterns in UX smells to broader hypotheses about UXDebt in IDEs using these criteria: 
\begin{itemize}[leftmargin=10pt]
    \item \textbf{Salience-Neglect Hypothesis:} A high density of smells associated with a highly valued IDE characteristic (e.g., efficiency) may signal that this dimension, despite being central to DEX, receives insufficient design attention, thus accumulating UXDebt.
    \item \textbf{Saturation-Resolution Hypothesis:} Conversely, if smells cluster around less valued characteristics (e.g., reliability), it may indicate that core experiential qualities (efficiency, intuitiveness) are relatively mature and that residual issues now emerge in peripheral dimensions.
\end{itemize}

\section{Results}

\subsection{UX Smells and IDE Qualities}

To ground the interpretive analysis, we first examined how the UX smell framework aligns with desirable IDE qualities identified by \cite{Kuusinen2015}. This mapping enables a dual perspective: highlighting which forms of UX friction are most frequent and revealing how the taxonomy itself resonates with the experiential expectations developers hold for their work environments. 
The distribution shows a clear concentration of UX smells around the cognitive-perceptual qualities \textbf{informativeness (6 UX smells), clarity (6), intuitiveness (5), efficiency (4), and ease of use (4)}), indicating that the framework primarily captures breakdowns in perception, comprehension, and control. Conversely, qualities such as \textit{flexibility}, \textit{empowerment}, and \textit{learnability} appear only marginally represented, suggesting that current UX smell taxonomies tend to diagnose short-term interaction breakdowns more than long-term experiential frictions.

To interpret these tendencies in relation to the broader model of DEX, we organized Kuusinen’s qualities into four higher-order clusters reflecting distinct experiential tensions: \textbf{cognitive transparency}, \textbf{flow efficiency}, \textbf{structural reliability}, and \textbf{peripheral experience}. 
Our four clusters refine (not extend) Fagerholm’s DEX framework \cite{fagerholm2012developer} by increasing analytic granularity: cognitive transparency maps to DEX’s cognitive dimension, flow efficiency spans cognitive/conative experience, and peripheral experience covers affective/conative facets. Structural reliability is made explicit as an enabling condition that, when degraded, systematically undermines cognitive and conative experience.

\subsection{Descriptive Statistics from VSCode Issues}

Out of 2350 analyzed issues, 61\% were identified as UX smells (1455 issues), while the remainder were bugs or feature requests.  

\begin{figure}
    \centering
    \includegraphics[width=.75\linewidth]{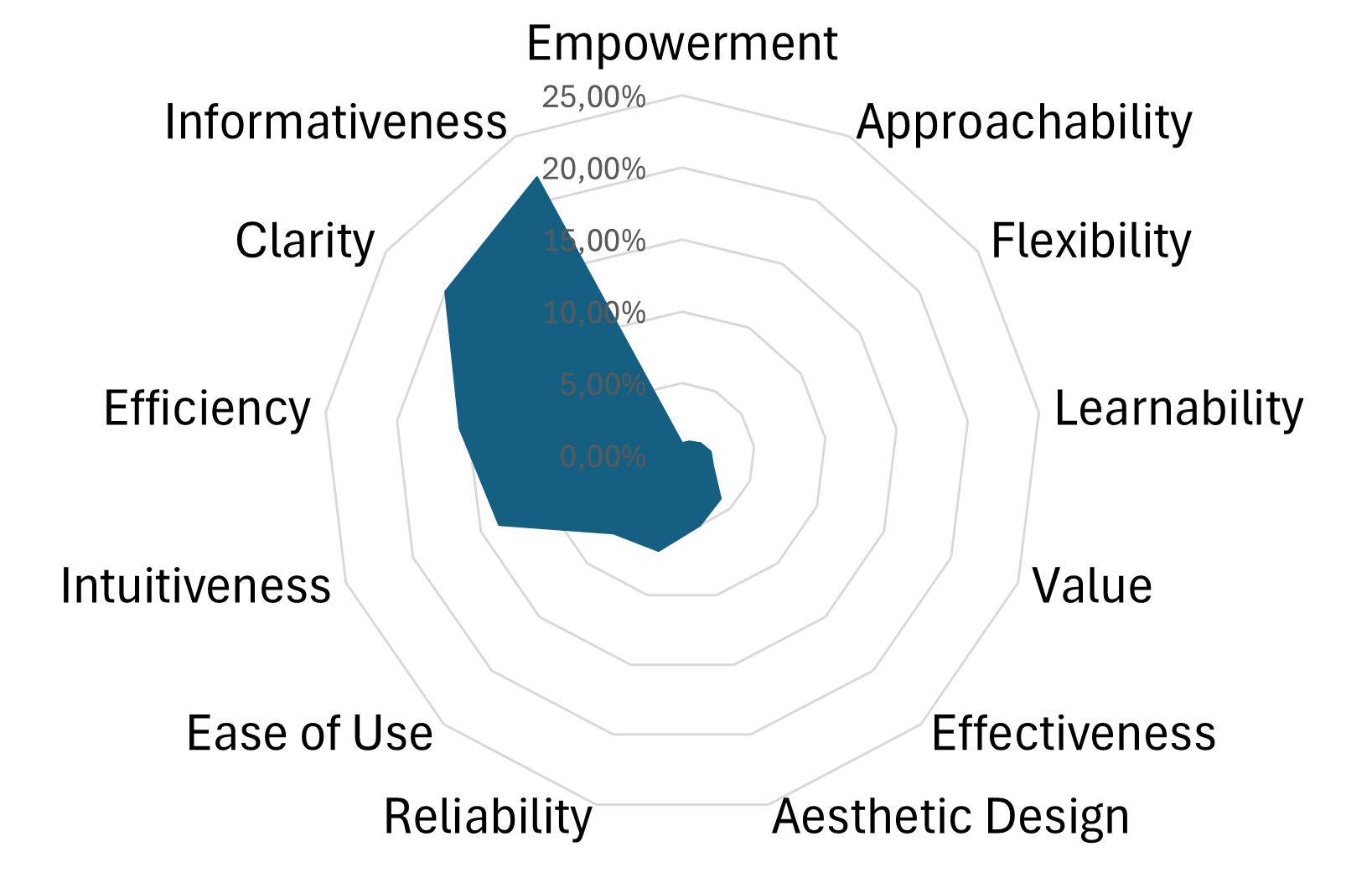}
    \caption{IDE profile according to UX friction}
    \label{fig:ideProfile}
    \Description{Figure 1 is a radar diagram that reflects the concentration of issues around IDE qualities, with a focus on Informativeness, Clarity and Efficiency}
\end{figure}

Among UX smells, mapping to Kuusinen’s 13 IDE quality dimensions yielded a markedly asymmetric distribution (see Table \ref{tab:uxsmells-kuusinen}). Over 70\% of selected issues cluster around \textit{Informativeness}, \textit{Clarity}, \textit{Efficiency}, and \textit{Intuitiveness}, the same dimensions Kuusinen identified as most valued by developers.  
Conversely, qualities linked to learning, autonomy, or flexibility represent less than 5\% of all cases, suggesting low visibility and prioritization in the design process (see Fig. \ref{fig:ideProfile}).

\newcommand{\ccell}[2]{\makebox[#1][c]{#2}}

\begin{table}[t]
\centering
\caption{UX smells across IDE Quality Dimensions}
\label{tab:uxsmells-kuusinen}
\footnotesize

\newlength{\ideqcolw}
\setlength{\ideqcolw}{2.35cm}
\newlength{\smellcolw}
\setlength{\smellcolw}{\dimexpr\columnwidth-\ideqcolw-1.1cm-1.1cm-6\tabcolsep\relax}

\begin{tabular}{@{}p{\ideqcolw}p{1.1cm}p{1.1cm}p{\smellcolw}@{}}
\toprule
\textbf{IDE Quality} &
\ccell{1.1cm}{\textbf{\# Issues}} &
\ccell{1.1cm}{\textbf{\% Issues}} &
\textbf{Frequent smells} \\
\midrule
Informativeness & \ccell{1.1cm}{312} & \ccell{1.1cm}{21.4\%} & Undescriptive Element, Inconsistent Feedback, No Progress Indicator \\
Clarity         & \ccell{1.1cm}{286} & \ccell{1.1cm}{19.7\%} & Overlooked Content, Gral. UI Inconsist., Unformatted Input \\
Efficiency      & \ccell{1.1cm}{223} & \ccell{1.1cm}{15.3\%} & Overloaded Menus, Distant Content, Inconsistent Spacing \\
Intuitiveness   & \ccell{1.1cm}{194} & \ccell{1.1cm}{13.3\%} & Misleading Link, Wrong Default Value, Inconsistent Placement \\
Ease of Use     & \ccell{1.1cm}{103} & \ccell{1.1cm}{7.1\%}  & Late Validation, Abandoned Form, Overloaded Menus \\
Reliability     & \ccell{1.1cm}{97}  & \ccell{1.1cm}{6.7\%}  & Unresponsive Element, Inconsistent Theming \\
Aesthetic Design& \ccell{1.1cm}{72}  & \ccell{1.1cm}{4.9\%}  & Clipped/Overlapping UI, General UI Inconsistency \\
Effectiveness   & \ccell{1.1cm}{58}  & \ccell{1.1cm}{4.0\%}  & No Client Validation, Scarce Search Results \\
Value           & \ccell{1.1cm}{33}  & \ccell{1.1cm}{2.3\%}  & Useless Search Results, Poor Accessibility \\
Learnability    & \ccell{1.1cm}{29}  & \ccell{1.1cm}{2.0\%}  & Inconsistent Placement, Poor Discoverability \\
Flexibility     & \ccell{1.1cm}{21}  & \ccell{1.1cm}{1.4\%}  & Forced Bulk Action \\
Approachability & \ccell{1.1cm}{16}  & \ccell{1.1cm}{1.1\%}  & Poor Discoverability, Poor Accessibility \\
Empowerment     & \ccell{1.1cm}{11}  & \ccell{1.1cm}{0.8\%}  & Forced Bulk Action \\
\bottomrule
\end{tabular}
\end{table}

\setlength{\belowcaptionskip}{-10pt}
\setlength{\abovecaptionskip}{0pt}

\subsection{Analytical and Interpretive Mapping }

\begin{figure}
    \centering
    \includegraphics[width=1\linewidth]{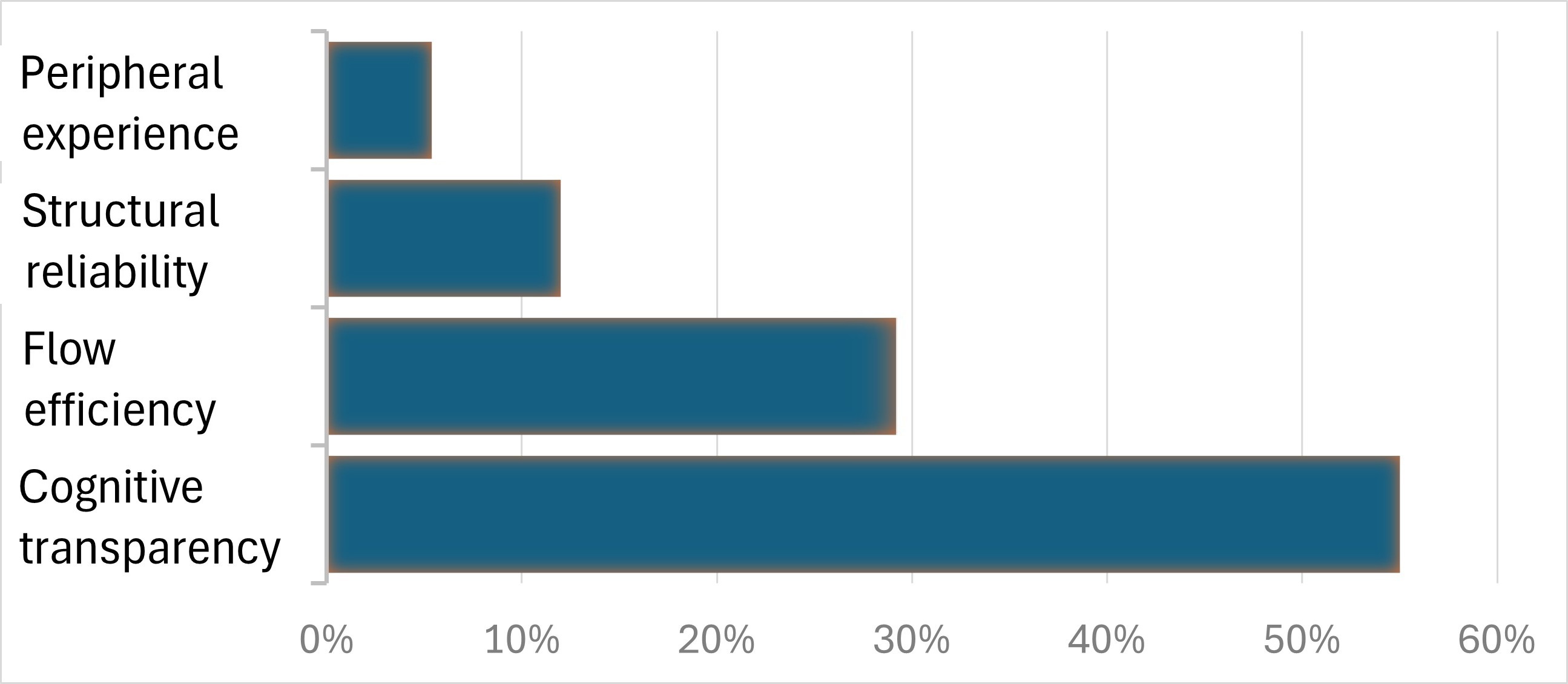}
    \caption{Bar graph with the UX smells dimensions clustering}
    \label{fig:clusters}
    \Description{Figure 2 is a bar graph reflecting the concentration of UX smells around the four experiential clusters. The highest bar corresponds to the Cognitive Transparency cluster, followed by the Flow Efficiency cluster, then Structural reliability cluster and the shortest corresponds to the Peripheral Experience cluster.}
\end{figure}

In this section, UX smells are examined within the four experiential clusters to trace how different types of UXDebt map onto the experiential qualities most valued by developers (see Fig. \ref{fig:clusters}).

\textbf{Cognitive Transparency Cluster (54\%).}
Comprising \textit{Informativeness}, \textit{Clarity}, and \textit{Intuitiveness}, this cluster concentrates over half of all UX smells.  
Frequent issues include insufficient feedback, unclear tooltips, ambiguous icons, and inconsistent visual cues, e.g. \textit{“The diff viewer shows changes but not which file is active”.}    
These cases reduce situational awareness and cognitive legibility, aligning with the \textit{cognitive dimension} of the DEX model \cite{fagerholm2012developer}.
UXDebt manifests as cognitive opacity, 
the accumulation of small inconsistencies and missing cues that erode the readability of system state over time.
\textbf{Flow Efficiency Cluster (29\%).}
Integrating \textit{Efficiency}, \textit{Ease of Use}, \textit{Value}, and \textit{Effectiveness}, this group covers issues that interrupt workflow continuity or require redundant steps, e.g.  
\textit{“Settings editor workspace folder selector dropdown opens too far away from tab”.}  
Such frictions reflect the disruption of the flow-related qualities highlighted by Kuusinen: efficiency and ease of use are central to how developers experience productivity within an IDE \cite{Kuusinen2015}. UXDebt in this cluster accumulates as process fragmentation, when local optimizations or feature additions compromise the seamless continuity of core workflows.
\textbf{Structural Reliability Cluster (12\%).}
Covering \textit{Reliability} and \textit{Aesthetic Design}, this cluster captures inconsistent feedback, delayed visual refreshes, and unsynchronized themes—e.g.,  \textit{“macOS: inconsistent UI when it comes to inputs border radius.”}  
This emerges as a form of structural UXDebt rooted in architectural or rendering constraints.
\textbf{Peripheral Experience Cluster (5\%).}
Encompassing \textit{Learnability}, \textit{Flexibility}, \textit{Approachability}, and \textit{Empowerment}, this cluster shows scarce representation. A few issues include discoverability or customization problems.  
Rather than indicating the absence of UXDebt, this low density may reflect latent or postponed debt in peripheral qualities, dimensions that receive less attention once functional stability is reached.

\setlength{\belowcaptionskip}{-10pt}
\setlength{\abovecaptionskip}{0pt}

These results, when examined in light of the hypotheses posited in Section \ref{sec:interpretiveAnalysis}, may be interpreted as follows:
\begin{itemize}[leftmargin=10pt]
    \item \textbf{Salience–Neglect Hypothesis}: we observed a high concentration of UX smells in \textit{Efficiency}, \textit{Informativeness}, and \textit{Clarity} (accounting for nearly 60\% of all cases).  
    This pattern suggests that the IDE’s most valued experiential qualities are also the ones most affected by UXDebt.  These core dimensions concentrate both functional complexity and user interaction, making them especially vulnerable to degradation through iterative growth and design trade-offs \cite{fagerholm2012developer}.
    We interpret this concentration as a bias in prioritizing functional expansion over cognitive experience: aspects that developers consider essential for productivity tend to accumulate subtle but pervasive usability frictions.
    \item \textbf{Saturation–Resolution Hypothesis}: is supported by the comparatively low frequency of issues in \textit{Reliability}, \textit{Learnability}, and \textit{Flexibility}.  Such scarcity may indicate functional maturity: once core mechanics and workflows stabilize, residual UX smells emerge primarily in peripheral or supporting dimensions, where design iteration is slower or less visible to users. This pattern aligns with the idea that UXDebt shifts from central to marginal layers as the product evolves.
\end{itemize}

Together, these observations highlight how UXDebt in VSCode evolves not merely through accumulation but through re-localization: from emergent friction in new features to persistent cognitive drag in long-standing ones.

\section{Conclusions and future work}

Our results show that VSCode exhibits a maturity pattern typical of large IDEs: most UXDebt concentrates in \textit{clarity}, \textit{informativeness}, and \textit{intuitiveness}, dimensions mediating the dialogue between interface and user, rather than technical reliability.  
This supports the view that in complex environments, UXDebt accumulates where interaction is most cognitive and frequent, not where code is most fragile.  
From a longitudinal perspective, this shift reflects an evolution from \textit{how it works} toward \textit{how it communicates and feels}.  
This study focuses exclusively on the core VSCode IDE to establish a baseline of UXDebt within the primary host platform. This may limit generalizability, as mature developers rely on a vast extension ecosystem. While the host’s architecture constrains UI disruption, UX friction could emerge from unforeseen extension interactions. 
Moreover, our reliance on GitHub issues only captures "reported" friction, potentially omitting "silent" UX smells.
Our future work will combine other data collection methods to mitigate these limitations. We also plan to compare our findings across different IDE ecosystems to determine if the concentration of UXDebt in cognitive transparency is an intrinsic feature of mature development environments.

\begin{acks}
Authors acknowledge grant PICT-2019-02485 from Agencia I+D+i.
\end{acks}

\bibliographystyle{ACM-Reference-Format}
\bibliography{vscode}

\end{document}